\documentclass[10pt, conference]{IEEEtran}

\usepackage{cite}
\usepackage[utf8]{inputenc}
\usepackage{booktabs}
\usepackage{balance}
\usepackage{comment}
\usepackage[pdftex]{graphicx}
\usepackage{amsmath}
\usepackage{array}
\usepackage{dblfloatfix}
\usepackage{url}
\usepackage{subfigure}
\usepackage{comment}
\usepackage[T1]{fontenc}
\usepackage{caption}
\usepackage{color}
\usepackage{amsmath}
\usepackage{amssymb}
\usepackage{amscd}
\usepackage{listings}
\usepackage{algorithm}
\usepackage{placeins}
\usepackage{float}
\usepackage[noend]{algorithmic}
\usepackage{paralist}
\usepackage{lstCypher}
\usepackage{paralist}

\begin{document}

\title{\emph{Portinari}: A Data Exploration Tool to Personalize Cervical Cancer Screening}

\author{
	\IEEEauthorblockN{
		Sagar Sen\IEEEauthorrefmark{1},
		Manoel Horta Ribeiro\IEEEauthorrefmark{2},
		Raquel C. de Melo Minardi\IEEEauthorrefmark{2}, 
		Wagner Meira Jr.\IEEEauthorrefmark{2} and
		Mari Nygård\IEEEauthorrefmark{3}
	}
	
	\IEEEauthorblockA{
		\IEEEauthorrefmark{1}
		Certus V\&V Center and HPV Research Group, \\ 
		Simula Research Laboratory and Cancer Registry of Norway, Oslo, Norway, \\
		Email: sagar@simula.no
	}
	
	\IEEEauthorblockA{
		\IEEEauthorrefmark{2}
		Department of Computer Science,	\\
		Universidade Federal de Minas Gerais, Belo Horizonte, Brazil, \\
		Email: \{manoelribeiro, meira, raquelcm\}@dcc.ufmg.br
	}
	
	\IEEEauthorblockA{
		\IEEEauthorrefmark{3}
		HPV Research Group,\\
		Cancer Registry of Norway, Oslo, Norway, \\
		Email: mari.nygard@kreftregisteret.no
	}
}

\maketitle

\begin{abstract}
Socio-technical systems play an important role in public health screening programs to prevent cancer. Cervical cancer incidence  has significantly decreased in countries that developed systems for organized screening engaging medical practitioners, laboratories and patients. The  system  automatically identifies individuals at risk of developing the disease and invites them for a screening exam or a follow-up exam conducted by medical professionals. A \emph{triage algorithm} in the system aims to reduce unnecessary screening exams for individuals at low-risk while detecting and treating individuals at high-risk. Despite the general success of screening,  the triage algorithm is a one-size-fits all approach that is not personalized to a patient. This can easily be observed in historical data from screening exams. Often patients rely on personal factors to determine that they are either at high risk or not at risk at all and take action at their own discretion. Can exploring patient trajectories help hypothesize personal factors leading to their decisions?  We present Portinari, a data exploration tool to query and visualize future trajectories of patients who have undergone a specific sequence of screening exams. The web-based tool contains (a) a visual query interface (b) a backend graph database of events in patients' lives (c) trajectory visualization using sankey diagrams. We use Portinari to explore diverse trajectories of patients following the Norwegian triage algorithm. The trajectories demonstrated variable degrees of adherence to the triage algorithm and allowed epidemiologists to hypothesize about the possible causes.
\end{abstract}


\IEEEpeerreviewmaketitle

\section{Introduction}
\label{sec:intro}

Software systems in several countries have been deployed to facilitate the prevention of cancer in society. For instance, at the Cancer Registry of Norway, a software system guides eligible women in the age group between 25 and 69 years to attend cervical cancer screening exams. They take tests commonly known as the \emph{Pap smear} at three year intervals or go through follow-up exams in  case of a higher risk. Women are  sent invitation  and reminder letters to their postbox or via \emph{digipost}\footnote{\url{https://www.digipost.no/}}. Many women  attend screening exams and follow up exams and this has reduced deaths due to cervical cancer by up to 80\%~\cite{CervicalScreeningEU2}. Outcomes of all screening exams are sent back to the cancer registry and registered in a relational database. This feedback loop of the software system with the society manifests itself into a \emph{socio-technical system}. The reliability of the socio-technical system greatly depends not only on how bugfree the software system is but also on how well we are able to use it to mobilize and engage people (medical practitioners, nurses, epidemiologists, and patients in the general public) into taking action against cervical cancer while keeping health expenditure to a minimum.

\begin{figure*}[!ht]
\centering
\includegraphics[width=0.8\textwidth]{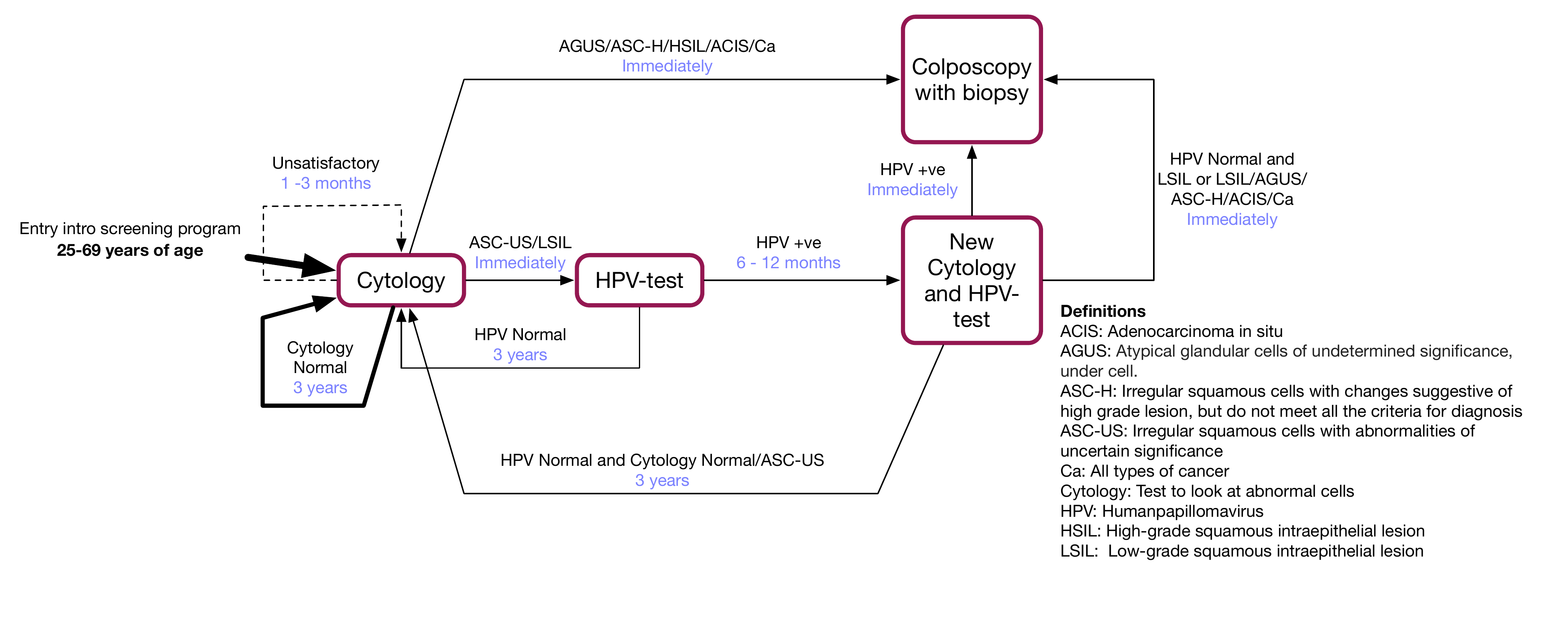}
\caption{Triage algorithm for cervical cancer screening in Norway.}
\label{fig:triage}
\end{figure*}

The main challenge of a socio-technical system for cervical cancer screening is to \emph{simultaneously}: 
\textit{(a)} \emph{Minimize over-screening of patients} due to its possible harmful effects \cite{bernie1998systematic} \cite{nobbenhuis1999relation} 
\textit{(b)} \emph{Increase screening attendance of under-screened patients} to reduce cancer burden on the state or costs incurred from cancer treatment~\cite{CostEffectivenessColorectalCancer}~\cite{CostEffectivenessCervicalCancer}.
A \emph{triage algorithm} on the population aims to optimize screening by inviting a patient to attend a screening exam ~\cite{CancerScreeningUS}~\cite{CervicalScreeningEU}. The triage algorithm  is implemented to automatically identify eligible patients in the population registry of a country and send invitations and reminders to them for screening. For instance, the triage algorithm in Norway\footnote{\url{https://www.kreftregisteret.no/globalassets/gammelt/cervix/flytskjema.pdf}}, shown in  Figure \ref{fig:triage}, automatically invites  women who turn 25 or immigrate into Norway at a later age to a cytology exam. The algorithm also invites women to a follow-up considering the outcome of the previous exam of a patient as detected in the anonymized cervical cancer screening database at the Cancer Registry of Norway. For instance, if a woman is tested with cytology ASC-US/LSIL she is immediately invited to take a test for the \emph{humanpappilomavirus} (HPV). This is because most cases of cervical cancer are due to a persistent infection of a high-risk variant of the \textit{human papillomavirus} (HPV)~\cite{HPVCC}. If the HPV test turns out to be positive, the patient is invited to a new cytology and HPV test in 6 to 12 months. This is because the human immune system can often eliminate the virus on its own. If the HPV infection persists the patient undergoes a colposcopy which can have harmful effects due to the biopsy. The screening guidelines have been very successful, reducing the incidence of cervical cancer by 70\%\cite{lonnberg2015cervical}.  The number of cervical cancer cases per year is around 300 instead of an estimated 1000 if the screening program did not exist.

Despite the general success of the screening program we observe in an anonymized cervical cancer screening database that people take or do not take tests based on their perceived risk of being diagnosed with cervical cancer. Low risk may be perceived due to many personal factors such as  a healthy lifestyle and  taking precautions against contamination. Fear and additional symptoms (which could be bleeding) not detected by screening exams can contribute to the perception of higher risk. Those who perceive their risk to be high may take too many tests which may have harmful effects. Those who perceive their risk to be low perhaps due to lifestyle factors may take very few or no screening exams at all. Therefore we ask, \emph{can exploring patient trajectories give us clues towards personalizing cervical cancer screening?} We believe that software tools to explore peoples' trajectories can help improve the social objective in a socio-technical system. In our case, the social objective amounts to personalizing cervical cancer screening and consequently reducing number of cancer cases.

\begin{figure*}[!ht]
\centering
\includegraphics[width=0.8\textwidth]{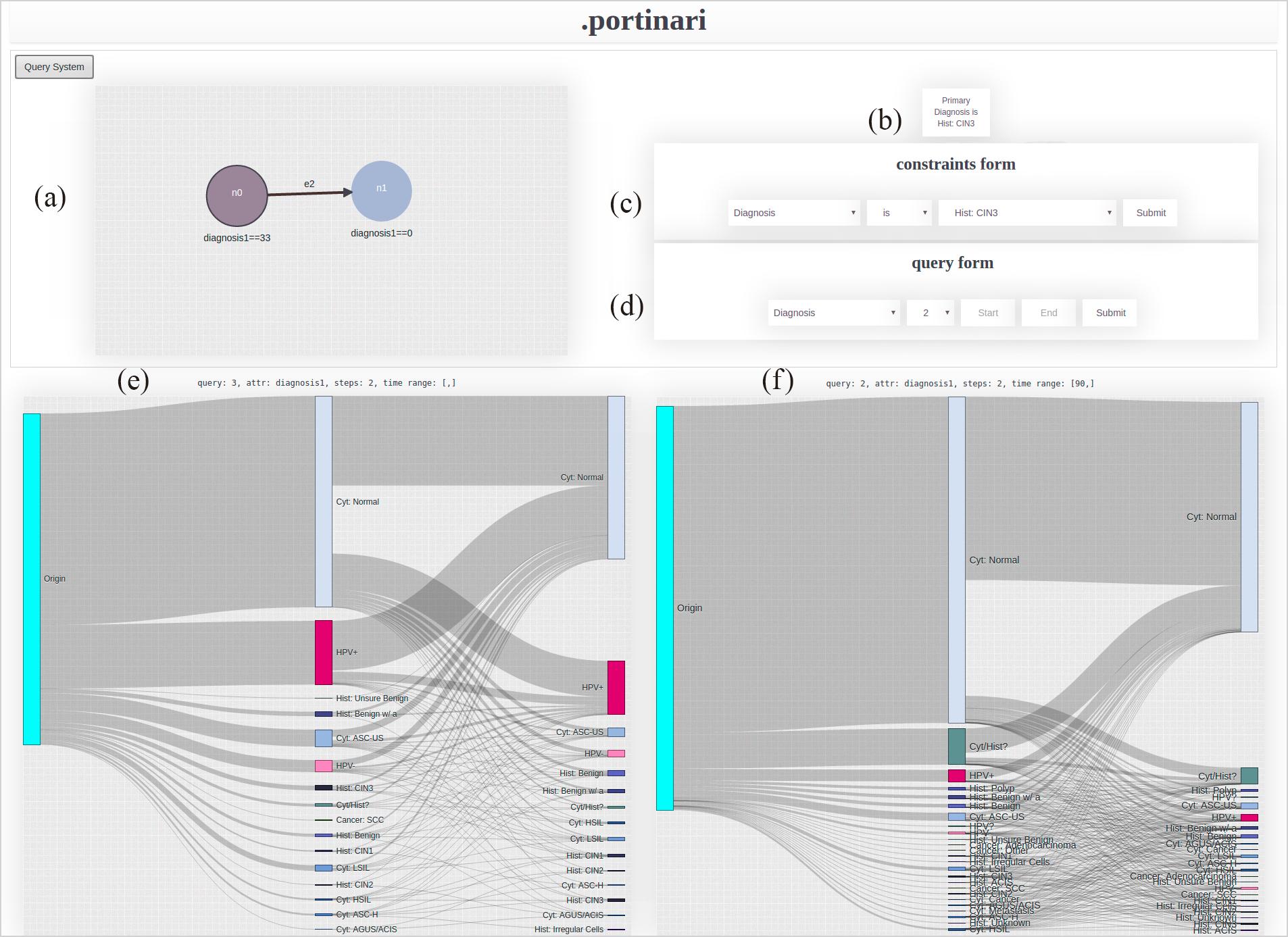}
\caption{\textit{Portinari}'s overall interface: \textit{(a)} A drag and drop canvas to create a query graph where nodes represent \emph{events} and edges representing \emph{order of events} \textit{(b)} Constraints created on properties of nodes and edges in the query graph \textit{(c)} Constraint creation form to choose property and specify a constraint \textit{(d)} A query form to specify a property to observe in events (e.g. diagnosis) of trajectories, number of events to show in the trajectory (e.g. 2) for patients satisfying the query graph \textit{(e)},\textit{(f)} Upto to two interactive Sankey diagrams displaying trajectories for the chosen event property (e.g. diagnosis).}
\label{fig:portinari}
\end{figure*}

\noindent \textbf{Approach:} To personalize cervical cancer screening, we exploit the screening outcomes of patients over a period of 20 years to understand if screening can be personalized. We achieve this in two steps:
\begin{enumerate}
\item Data from a socio-technical system for cervical cancer is  available in the form of events in the life of patients. In cervical cancer screening an event corresponds to attendance to an exam such as a HPV test along with date and type of diagnosis. We transform these events from its flat form of transaction records into sequences of connected events for individual patients in a \emph{graph database} implemented in Neo4J \cite{developers2012neo4j}.

\item The main contribution of the paper is a web-based data exploration tool \emph{Portinari}\footnote{Available upon request}, to explore and visualize individual\footnote{We use the terms \emph{patient} and \emph{individual} interchangeably in the article.} trajectories by querying the graph database. The interface of the tool at a glance is shown in Figure~\ref{fig:portinari}. The interface allows users (epidemiologist specializing in cervical cancer screening) to visually specify a \emph{query graph} representing a sequence of exams and respective diagnoses taken by an individual. The user may also specify constraints such as the time between any two diagnosis and/or age ranges. \emph{Portinari} automatically generates future trajectories of patients who underwent the input sequence of exams and diagnosis by matching similar patients in the graph database. \emph{Portinari} visualizes the outcome as a sankey diagram~\cite{martin2012visualizing} of a patient trajectory for a finite number of subsequent steps using the query subgraph as the origin. Sankey diagrams are a specific type of flow diagram, in which the width of the arrows is shown proportionally to the flow quantity. The flow quantity in this paper typically represents number of individuals going from one exam to another.
\end{enumerate}

We use \textit{Portinari} to explore a graph database of screening exams from the Norwegian Cervical Cancer Screening program. The  large anonymized database contains high quality data events from about $906713$ distinct women in Norway, such as the date and results of various types of exams related to cervical cancer screening~\cite{NorwegianCancerRegistry1}~\cite{NorwegianCancerRegistry2}.  \textit{Portinari} was able to assess  many of the recommendations of the Norwegian triage algorithm, and also find trajectories where the guidelines could use some further personalization. For instance, an interesting finding was that there were several patients who returned within 30 days of having normal cytology. Few of them got a more advanced test such as the histology done while some others were diagnosed with HPV positive. This was in contrast to the triage algorithm's recommendation to return in three years. This result clearly shows that the cytology exam is not perfect and there are other symptoms such as bleeding that are not taken into account to personalize recommendations. Tools for data exploration such as \textit{Portinari} are a necessity to inspect and validate rapidly evolving socio-technical software systems, particularly in public health. Data is not static in these systems and is constantly updated by societal behavior. Policy makers should consider outliers and personalized scenarios to update their triage algorithms to enhance prevention.

The rest of the paper is organized as follows. In Section~\ref{sec:graph}, we present the graph database representation of the events from screening exams for a population. The data exploration system, \textit{Portinari}, is described in Section~\ref{sec:explore}.
We made explorations inspired by the Norwegian triage algorithm using \textit{Portinari} and the results and its implications are presented in Section~\ref{sec:explorations}. Related work is presented in Section~\ref{sec:related}. Finally, we present conclusions and future work in Section \ref{sec:conclusion}.

\section{Graph Database of Screening Events}
\label{sec:graph}

Events from societal behavior, such as people attending exams, are often stored by information systems as unrelated records of transactions. In cervical cancer screening, a screening exam and its diagnosis for a patient is such a record in a relational database. Creating temporal queries in relational databases is complex~\cite{COQUITO}~\cite{TQUEL}. Therefore, we advocate the representation of the temporal ordering using a graph database.

A patient's trajectory is in fact is these records put together in a chronological sequence. This is ideally modeled as a \emph{directed property graph} where nodes represent an exam event along with properties such as diagnosis type and stage of disease and edges represent the \emph{sequence between exams} with properties such as time between exams. A graph database representation of exam records allows querying and reasoning about patient trajectories as a whole. Below we describe both a tabular representation of the patients attending a screening exam and the transformation to a graph database representation. 

\noindent \textbf{Tabular Representation:} The cervical cancer screening  database in Norway contains high quality  events related to cervical cancer screening exams, dates and results of screening and diagnostic tests performed, as well as information about treatment of pre-cancers. It is estimated to be close to 100\% complete as it is mandatory by law to report all screening tests, cancers, emigration and deaths~\cite{NorwegianCancerRegistry1}~\cite{NorwegianCancerRegistry2}. The database contains about records of 5 million exams taken by 0.9 million women from the year 1992 to 2014. Each record in the database is specified by a set of fields as presented in Table \ref{tab:record}. We anonymize the original database before running all the experiments in this article.  The anonymization was done by replacing all dates (birth date, diagnosis date, and censor date) to the 15th of the month, and perturbing every month by fuzzy factor of number between +/- 4. The resulting database had a very low re-identification risk as evaluated by the anonymization tool ARX\cite{prasser2014arx} while preserving the general characteristics of screening behaviour.

\begin{table}[]
\centering
\caption{Fields in the cervical cancer screening database.}
\label{tab:record}
\begin{tabular}{@{}ll@{}}
\toprule
\multicolumn{1}{c}{ID}             & \multicolumn{1}{c}{Internal numeric identifier.}     \\ \midrule
\multicolumn{1}{l|}{birthdate}     & Month and year of birth of the woman.                \\
\multicolumn{1}{l|}{diagnosisdate} & Date of diagnosis.                                   \\
\multicolumn{1}{l|}{type}          & Type of visit/exam.                                  \\
\multicolumn{1}{l|}{diagnosis}     & Result given by the exam/visit.                      \\
\multicolumn{1}{l|}{stage}         & Stage for cancer diagnosis.                          \\
\multicolumn{1}{l|}{lab\_nr}       & Code for the laboratory, which is a 5-digit number.  \\
\multicolumn{1}{l|}{region}        & Norwegian health care region.                        \\
\multicolumn{1}{l|}{censordate}    & Date for emigration/death/cervical cancer diagnosis. \\ \bottomrule
\end{tabular}

\end{table}
	
\begin{figure}[t]
\includegraphics[width=0.8\linewidth]{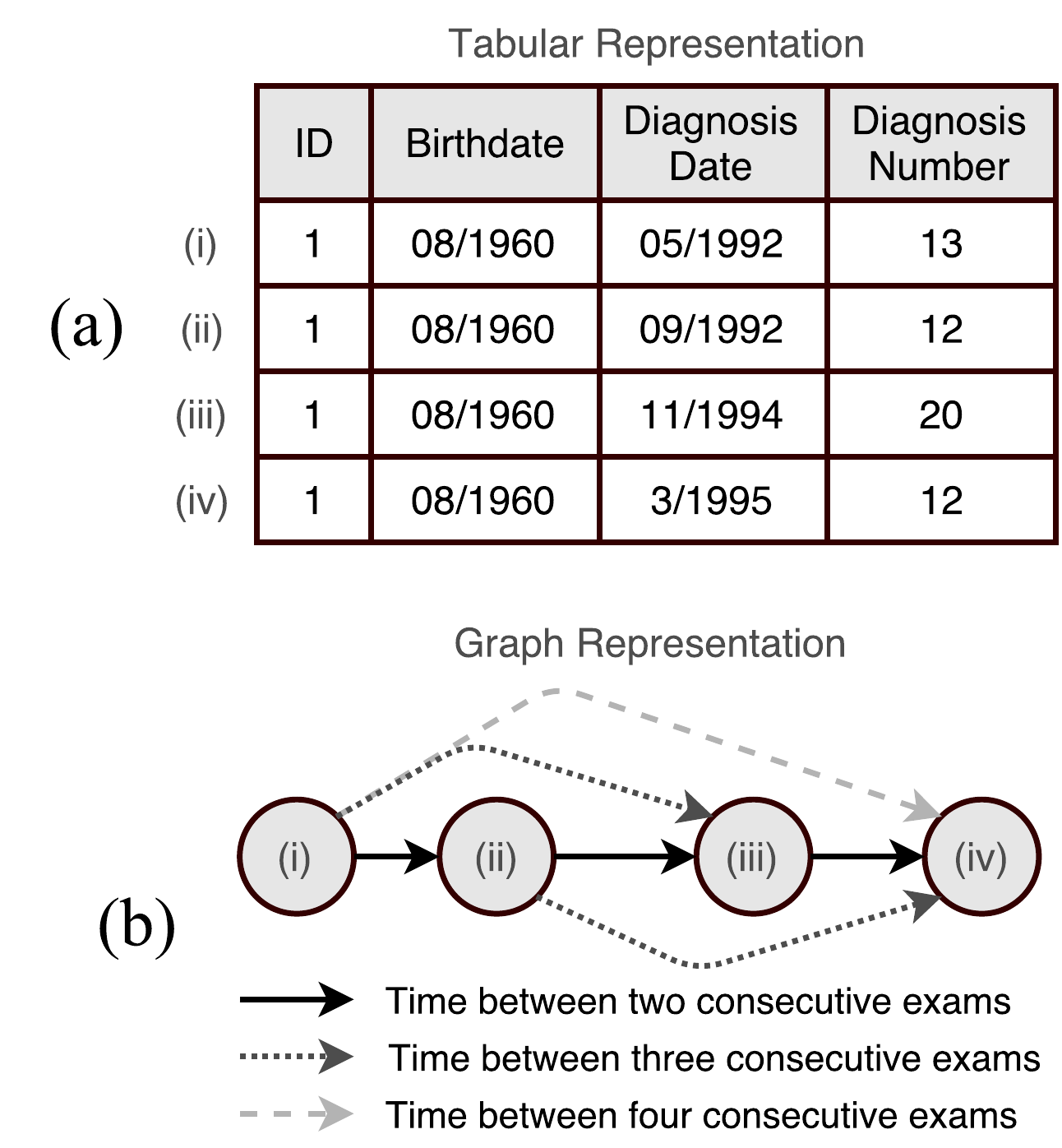}
\caption{Transformation of \textit{(a)} tabular data for a patient with a given ID to \textit{(b)} graph for the patient.}
\label{fig:dbtransf}
\end{figure}

\noindent \textbf{Graph Database Representation:} Anonymized tabular data of exam records for each patient is transformed to a graph database as illustrated in Figure~\ref{fig:dbtransf}. Exam records are processed in chronological sequence to create nodes representing \emph{events} such as exams with properties such as diagnosis type, stage, laboratory number. The edges, labeled \emph{next}, contain the time elapsed between two events stored as a property. In this paper we specify time in \emph{number of days}. We create additional edges between the nodes that are not consecutive but two or more events in the future as shown in Figure~\ref{fig:dbtransf}. This inclusion of edges between hoping over  nodes in time facilitates creation and faster execution of temporal queries between two non-consecutive events. For instance, a query that asks to return all patients who got cancer in 400 days from their first cytology positive simply will require matching a \emph{next} edge with number of days $<400$ days. This saves computation time by avoiding matching  all events and edges.

Every patient's trajectory is created and stored as a graph in the database implemented in Neo4J. An example Cypher statement to create a patient's trajectory is shown in Listing~\ref{listing:graphExample}. The statement is
generated from the table for the individual shown in Figure \ref{fig:dbtransf}\textit{(a)}. We first create nodes $p1$, $p2$, $p3$, and $p4$ that correspond to the exams taken by the patient, then the \emph{1-hop} edges with the time between consecutive exams, then the \emph{2-hop} edges and the \emph{3-hop} edge. The database does not contain relationships between patients and hence the set of subgraphs of patients are unconnected. In the future, if relationships such as friendship or family are made available by approval from an ethical committee we can  easily create edges in the schema-less graph database with this information. This is the main advantage of using  a graph database. Similarly, genetic information from patient's blood serum in the Janus bank \cite{jellum1995experiences} of the Cancer Registry of Norway and lifestyle factors such as smoking, alcohol consumption obtained from a survey \cite{hansen2011factors} can be introduced into the graph database to enrich it for more insight into societal processes.

\begin{lstlisting}[language=cypher, basicstyle=\tiny,
 frame=single, framerule=0.2pt, captionpos=b,
caption={An example Cypher query generated to create a patient trajectory in a graph database.}, label={listing:graphExample}]
CREATE 
(p1:Event // Create first node
	{PatientID:1, DiagnosisDate:"15.05.1992", BirthDate:"12.08.1960", ExamType:"cyt", 
	 Diagnosis:13, MorphologyCode:76700, CDate:NULL, LaboratoryNbr:19, Region:2}),
(p2:Event // Create second node
	{PatientID:1, DiagnosisDate:"15.09.1992", BirthDate:"12.08.1960", ExamType:"cyt", 
	Diagnosis:12, MorphologyCode:69000, CDate:NULL, LaboratoryNbr:19, Region:2}),
(p3:Event // Create third node
	{PatientID:1, DiagnosisDate:"15.11.1994", BirthDate:"12.08.1960", ExamType:"cyt", 
	Diagnosis:20, MorphologyCode:76700, CDate:NULL, LaboratoryNbr:19, Region:2}),
(p4:Event // Create third node
	{PatientID:1, DiagnosisDate:"15.03.1995", BirthDate:"12.08.1960", ExamType:"cyt", 
	Diagnosis:12, MorphologyCode:76700, CDate:NULL, LaboratoryNbr:19, Region:2}),
// 1-hop edges
(p1)-[:Next1 {SinceLast:123}]->(p2),
(p2)-[:Next1 {SinceLast:791}]->(p3),
(p3)-[:Next1 {SinceLast:116}]->(p4), 
// 2-hop edges
(p1)-[:Next2 {SinceLast:914}]->(p3), (p2)-[:Next2 {SinceLast:807}]->(p4),
// 3-hop edges
(p1)-[:Next3 {SinceLast:1030}]->(p4)
\end{lstlisting}

\section{Data Exploration System}
\label{sec:explore}

\textit{Portinari} is a data exploration system with the aim of exploring and possibly validating how software systems impact  social behavior. Socio-technical systems can comprise of software systems in a feedback loop with the society where data about peoples' behavior is gathered over time. \textit{Portinari} is a tool that allows inspection, exploration, and visualization of peoples' trajectories in a socio-technical system. It is a generic tool that can be used to query and visualize trajectories of events in any socio-technical  system. Data in socio-technical  systems evolve over time and hence need to be constantly explored to gain insight into the effectiveness of an organized social process such as the triage algorithm in a cervical cancer screening program. \textit{Portinari} was developed in light of such online monitoring of societal behavior.  In this article, we use \textit{Portinari} to explore trajectories of patients in the Norwegian cervical cancer screening  program. In the following sub-sections, we present an overview of \textit{Portinari} and describe the two principal steps in using the tool.

\subsection{System Overview}
\label{sec:so}

\textit{Portinari} relies on a  graph database of individuals' trajectories. For instance, we use the graph database of people attending to cervical cancer screening as described earlier in Section \ref{sec:graph}. It expects the graph database to contain nodes representing events and directed edges representing the chronological order between between events for each individual. The graph database is a collection of directed graphs of people's trajectories. The graph database is queried by \textit{Portinari}'s web-based interface as shown in Figure~\ref{fig:portinari}. The interface contains two parts:

\noindent \textbf{Querying Interface:} \textit{Portinari} has a web-based  querying interface (Figure~\ref{fig:portinari}\textit{(a-d)}) that permits modeling a \emph{query graph} representing a partial trajectory of an individual along with constraints. The length of the partial trajectory has an upper bound of the maximum number of exams a woman has ever taken (54 at present). Epidemiologists, explore finite sequences from the triage algorithm graph limited to two or three nodes representing exams. The query graph is transformed into query in the language Cypher \cite{team2013cypher} (a language similar to SQL but for graph databases) that is executed on the backend  graph database. The result of the query is the set of all future trajectories of patients that matched the input partial trajectory. 

\noindent \textbf{Results Interface:} The result (Figure~\ref{fig:portinari}\textit{(e-f)}) is visualized as an interactive sankey diagram \cite{riehmann2005interactive}. Sankey diagrams are a specific type of flow diagram, in which the width of the arrows is shown proportionally to the flow quantity. We use sankey diagrams to show how many people flow from an event to other events. \textit{Portinari} allows visualization of two sankey diagrams simultaneously, enabling the user to compare two scenarios.

\subsection{Creating a Query Graph}
\label{sec:qi}

\begin{figure}[t]
	\centering
\includegraphics[width=0.7\linewidth]{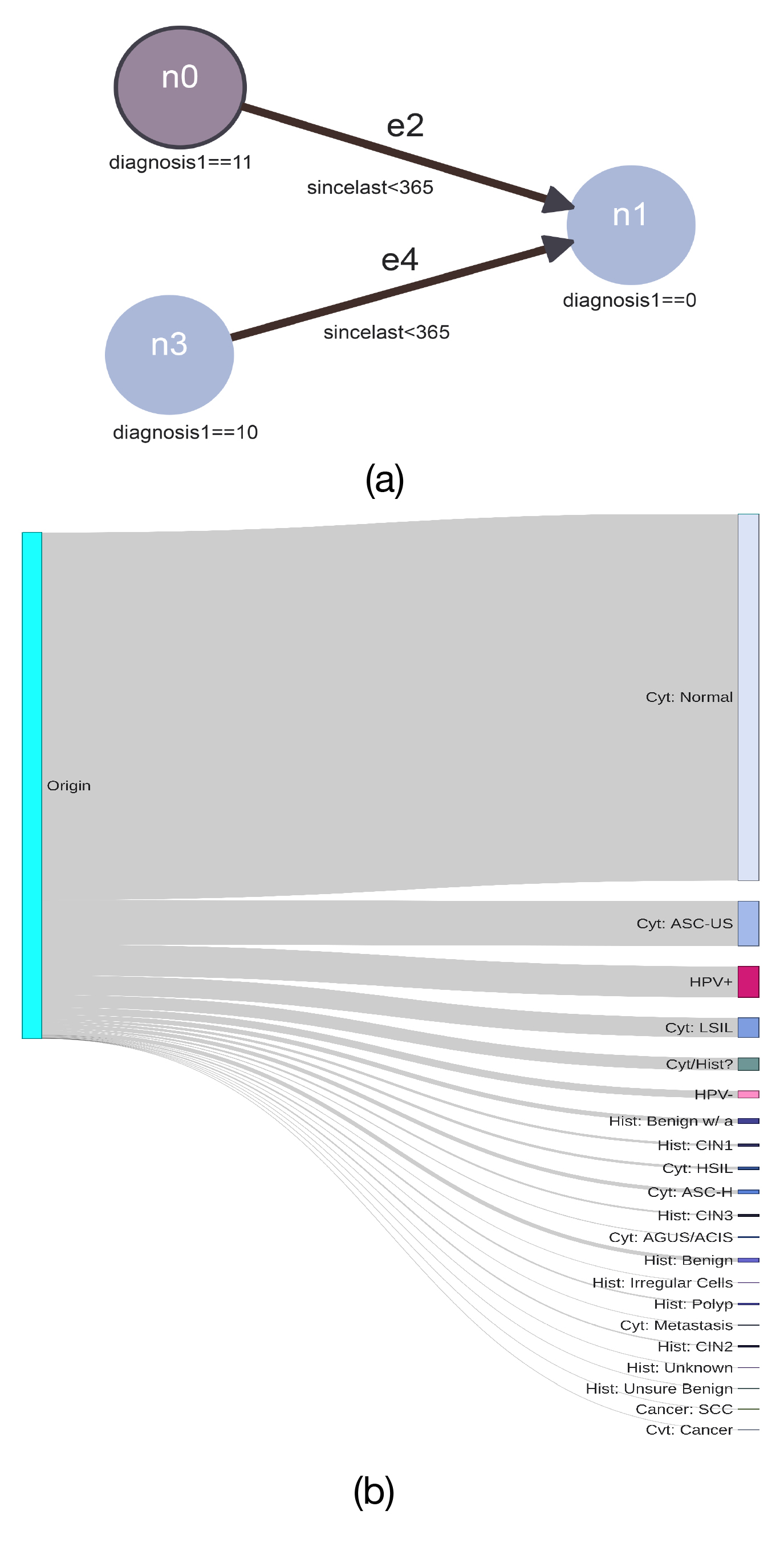}
\caption{ \textit{(a)} A query graph in \textit{Portinari} is visually specified where nodes represent events and edges represent order between events. Constraints on nodes and edges can be specified and appear below them \textit{(b)} \textit{Portinari} generates a sankey diagram in its results interface, displaying the trajectory of all individual that satisfy the graph query.}
\label{fig:query-int}
\end{figure}

The first step in using \textit{Portinari} is to create a \emph{query graph}. \textit{Portinari}'s query mechanism  has a drag-and-drop interface where the user builds the query graph representing \emph{sequence of events} in a patient's life to be found in the graph database.  
The user may: 
\textit{(a)} Add new nodes representing events
\textit{(b)} Create edges between nodes representing the sequence of events
\textit{(c)} Delete nodes.
In the cervical cancer screening context,  for instance, a node is an exam taken by a patient and the diagnosis she received. The edges represent a chronological ordering between two exams.

\noindent \textbf{Constraints:} The query interface allows to add constraints such as "\textit{BIGGER THAN}", "\textit{SMALLER THAN}", "\textit{EQUALS TO}" and "\textit{NOT EQUALS TO}" to properties in both nodes and edges. A specific node will be matched in the graph database if it satisfies \emph{all the associated constraints} on the node. Constraints on edges allow for instance the specification of a time range between two events.
Constraints are specified in the \textit{constraints form} as shown earlier in Figure~\ref{fig:portinari}\textit{(c)}. The constraints on a node are automatically rendered in query graph right below nodes and edges. They are also shown on top  of the \textit{constraints form} (Figure~\ref{fig:portinari}\textit{(b)}).

\noindent \textbf{Multiple Partial Trajectories:} The user may create more than one partial trajectory in the query graph. The query system considers  partial trajectories such as in Figure ~\ref{fig:query-int}\textit{(a)} as patients that match both the trajectory $n0-e2-n1$ and $n3-e4-n1$. Here $n0$, $n1$, $n3$ are event nodes and $e2$, $e4$ are edges between events in the multiple path partial trajectory.

\noindent \textbf{Query form:} The query form is shown in Figure~\ref{fig:portinari}\textit{(d)}. In this final step  the user specifies additional parameters to generate a sankey diagram using the \textit{query graph} as input. The user may specify the following parameters:
\textit{(a)} Number of subsequent events the user wants to visualize considering all patients satisfying the query graph as origin.
\textit{(b)} The timeline of a chosen attribute the user wants to visualize in the trajectory rendered as the Sankey diagram.
\textit{(c)} The time range for next event considering the query graph as origin.

\noindent \textbf{Example of use:} Consider the  scenario where a user wants to visualize the diagnosis of patients that had a specific pattern of exams and returned for an exam within $60$ days.
Lets assume that the cohort of interest is composed of patients who were diagnosed \textit{Normal} or \textit{Unsatisfactory} in the cytology exam (codes $11$ and $10$ respectively), and then, in less than a year, were diagnosed with HPV positive (code $0$) in any region.  An user employs the drag-and-drop interface to create a query graph with two possible paths  and constraints as shown in Figure~\ref{fig:query-int}\textit{(a)}. The constraints on the variable diagnosis for the events are shown under the node. While the constraint on the time $<365$ days is shown under the edges. The Cypher query generated by \textit{Portinari} from the visual interface is shown in Listing \ref{listing:cypherExample}. The nodes $n0$, $n1$, $n3$ and edges $e2$, $e4$ in the query correspond
to the query graph in Figure \ref{fig:query-int}\textit{(a)}. The node $nf1$ and the edge $ef1$ are used as placeholders to obtain the first exam taken by after patient pattern specified in the query graph.

\begin{lstlisting}[language=cypher, basicstyle=\tiny,
 frame=single, framerule=0.2pt, captionpos=b,
caption={An Example Cypher query generated to create a patient trajectory in a graph database.}, label={listing:cypherExample}]
MATCH ( n0:Event { Diagnosis:11 }) 
MATCH ( n1:Event { Diagnosis:0  })
MATCH ( n3:Event { Diagnosis:10 })
MATCH (n0)-[e2]-(n1)
	WHERE e2.SinceLast < 365
MATCH (n1)-[e4]-(n3)
	WHERE e3.SinceLast < 365
MATCH (n3)-[ef1:Next1]->(nf1)
	WHERE ef1.SinceLast < 60
RETURN COUNT(distinct n3), nf1.Diagnosis 
\end{lstlisting}

Finally, the user specifies the number of steps in future and the attribute he/she wants to visualize in the \textit{query form}. The user also can specify the time range for the immediate next event for all patients matching the query graph. As mentioned previously, the user wants the next diagnosis for patients that returned in more than $60$ days.  Therefore, the user can specify this constraint in the \textit{query form} and submit the query to obtain the results. We present how the results are rendered in the following subsection.

\subsection{Generating Patient Trajectories as Sankey Diagrams}
\label{sec:ri}

\textit{Portinari} presents results as an \emph{interactive sankey diagram} of the \emph{subsequent patient trajectories} who match the query graph. The sankey diagram as shown in Figure \ref{fig:query-int}\textit{(b)} consists of \emph{vertical bars} and \emph{flows} between the bars.  The leftmost vertical bar in the sankey diagram is called the \textit{Origin}. It represents the total number of patients who match the  \textit{query graph} specified in the query interface. The subsequent bars in the sankey diagram  represent the number of patients for different values of a chosen property of an event. In Figure~\ref{fig:query-int}\textit{(b)}, the different values correspond to the property \emph{diagnosis} of the event \emph{screening exam}. We are typically interested in the trajectory of diagnosis for patients.
The sankey diagram is interactive. Hovering the mouse cursor over the nodes displays a tool-tip informing how many patients converge to that event and their percentage in the population. Flow width between bars in the sankey diagram is directly proportional to the number of patients that took this specific path.  Hovering the cursor over the flows displays a tool-tip informing the number of patients  that went from the event represented by the source of the flow to the event represented by the target of the flow. The tool-tip also displays what percentage of patients the flow represents in the source node. The width of the flow gives the user a perception of how likely people are to follow one path compared to another. For instance, the path to \textit{Cytology:Normal} is far more likely than the one to \textit{Cancer} in Figure~\ref{fig:query-int}\textit{(b)}. \textit{Portinari} can render up to two sankey diagrams allowing the comparison of two different query graphs. 

\section{Explorations}
\label{sec:explorations}

We hope that exploring the graph database of patient trajectories will give us insight into personalizing the cervical cancer screening program in Norway. We use \textit{Portinari} to address the following broad questions:

\noindent \textbf{Q1:} What happens to patients who follow the triage algorithm?

\noindent \textbf{Q2:} What happens to patients who \emph{do not follow} guidelines in the triage algorithm?

\noindent \textbf{Q3:} Can we compare different trajectories taken by patients in the triage algorithm?

\begin{figure*}[t]
\includegraphics[width=\linewidth]{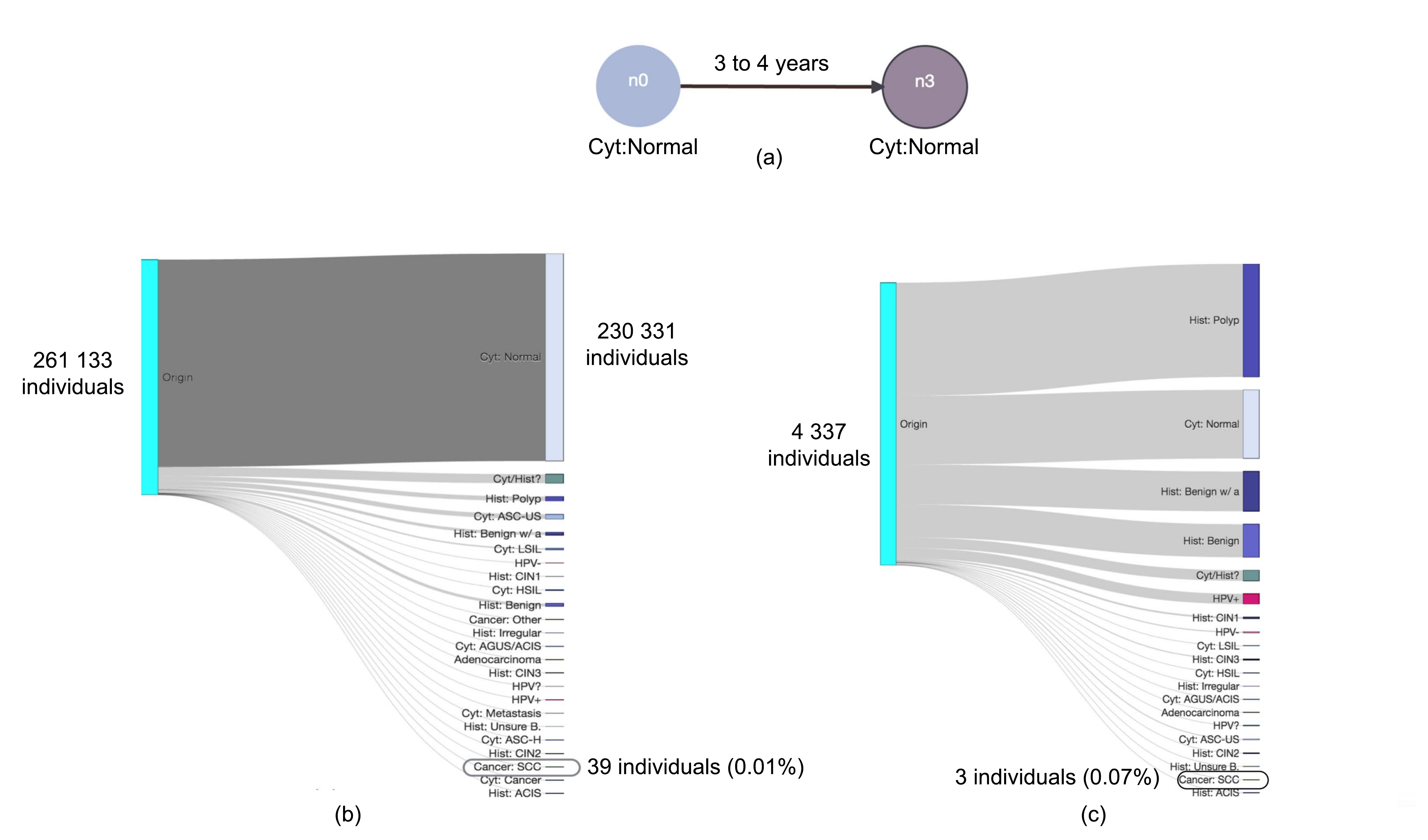}
\caption{ \textit{(a)}  Query graph for women with two consecutive normal cytologies \textit{(b)}  Trajectory of women who come back in 3 to 4 years after two normal cytology exams  \textit{(c)} Trajectory of women with two consecutive normal cytologies but return in just 30 days.}
\label{fig:cytNormal}
\end{figure*}

Question \textbf{Q1} is  essential to understand whether guidelines in a public health program are effective. In the triage algorithm shown earlier in Figure \ref{fig:triage}, the most commonly followed guideline is simple (shown in bold) for women to understand and follow. Women are asked to perform a cytology exam when they enter Norway and are above 25 years of age or when they turn 25. If the result of the test is normal then they are asked to come back every three years for regular exams. Therefore, we model the query graph for such a scenario in Figure \ref{fig:cytNormal}\textit{(a)}, where we are interested in the trajectories taken by patients who have two consecutive normal cytology exams. We use \textit{Portinari} to find out what happens to these women if they come back within three and four years after two normal cytologies. In Figure \ref{fig:cytNormal}, we see that about 88.2\% of the women were still diagnosed with cytology normal while some had more tests done. We also observe 39 women (about 0.01\%) who had a cancer (squamous cell carcinoma). These outliers despite following the triage algorithm, were still diagnosed with cancer. How could we have better screened these 39 women? Should they have undergone other tests in addition to the cytology exam? These are questions that exploration using \textit{Portinari} can raise.

Question \textbf{Q2} concerns women who do not adhere to the triage algorithm. In Figure \ref{fig:cytNormal}\textit{(c)}, we explore trajectories of women who return only 30 days after two consecutive normal trajectories. There were about 4337 women who returned only in 30 days, some of who got different tests such as HPV testing and histology. Some were diagnosed with HPV+ve while the histology revealed a \emph{polyp} in others. A polyp is an abnormal growth of tissue projecting from a mucous membrane. The most likely hypothesis is that the women or their doctors perceived a higher risk probably based on additional evidence such as bleeding that is not revealed by a cytology exam. These deviations from what the triage algorithm expects strengthens the case for personalization and improvements in testing and evidence collection.

\begin{figure}[t]
\includegraphics[width=\linewidth]{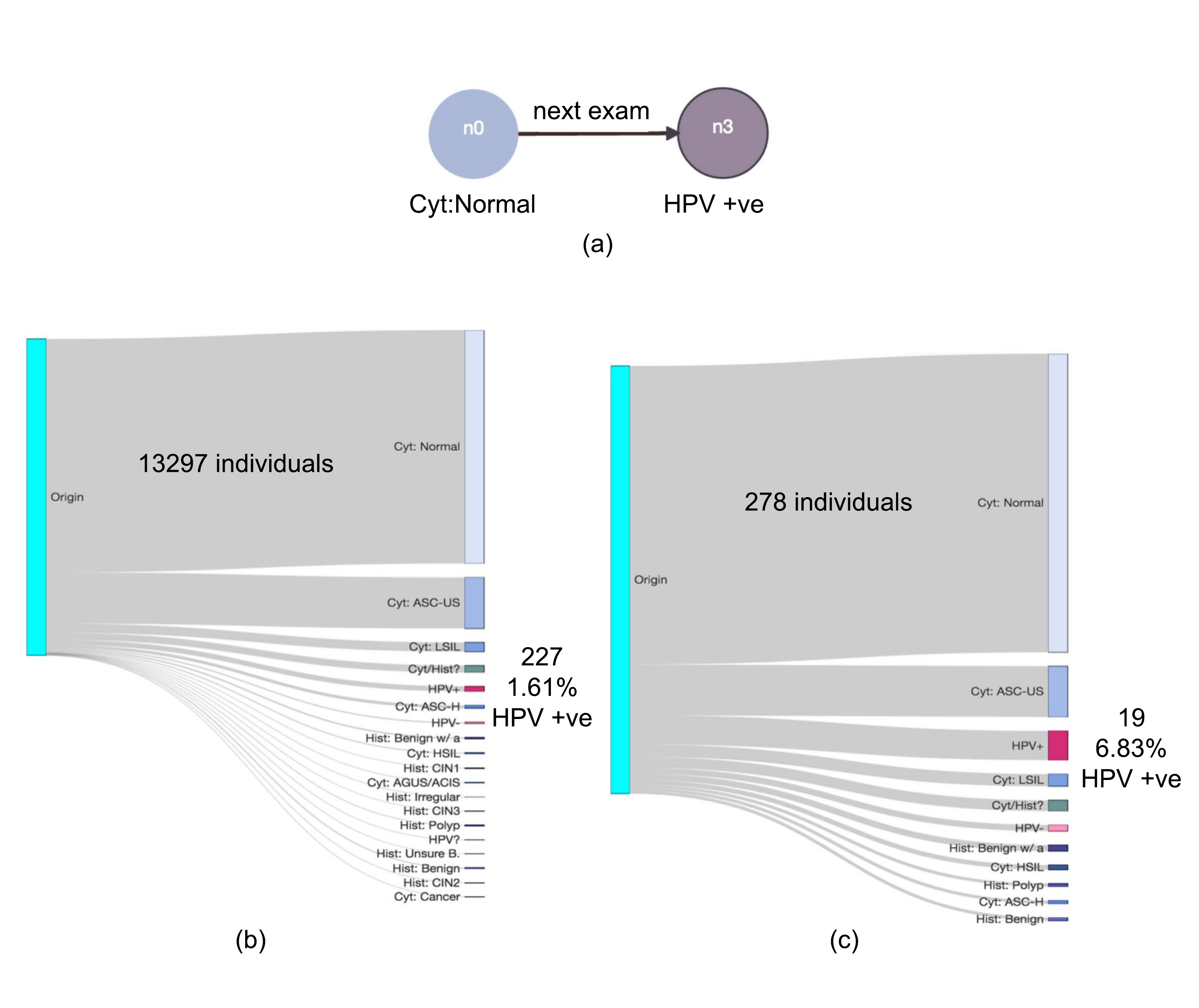}
\caption{\textit{(a)} Query graph for women who have a cytology normal and they take a HPV test and turn positive in the next exam \textit{(b)} Trajectory of women who come back in one year after the query graph \textit{(c)} Trajectory of women who come back after two years. }
\label{fig:compareTwo}
\end{figure}

Finally, for question \textbf{Q3}, we compare two future trajectories for one query graph. In Figure \ref{fig:compareTwo}, we construct a query graph illustrating the scenario where a cytology normal exam is  followed by a HPV exam that turns out to be positive. We compare trajectories for women who return within one year as recommended by the triage algorithm in Figure \ref{fig:compareTwo}\textit{(b)} and women who return between one and two years or who are delayed beyond 2 years in Figure \ref{fig:compareTwo}\textit{(c)}. We observe that majority of women that had HPV+ve test return within a year. This illustrates that they perceive their risk to be high when they have a positive test for the humanpapillomavirus. HPV sounds like HIV and does it somehow increase the gravity of the disease? Both trajectories show that few women have a persistent HPV infection. Which means that those who were diagnosed with HPV+ve first were not diagnosed a second time with HPV+ve. Many have a normal cytology diagnosis that follows the HPV+ve test. These trajectories explain the natural history of the HPV virus in the human body. Even the high risk variants of HPV  are  eliminated by the immune system in most cases. 
Comparison of two similar trajectories allows users to evaluate the trade-offs between coming on time as recommended by the triage algorithm or being delayed with respect to the screening algorithm. For policy makers and epidemiologists the comparison tool allows augmentation of the triage algorithm with personalized suggestions if necessary. For instance, people with weaker immune systems may need to be screened more frequently and asked to adjust their lifestyle  compared to those who are in good health over a threshold number of screening exams.

\section{Related Work}
\label{sec:related}
\textit{Portinari} is a generic exploration tool for socio-technical systems which we use to facilitate prevention of cancer based on interactive visualization of the societal process of cervical cancer screening. It derives inspiration from prior work on interactive tools for visualizing electronic health records (EHR) and querying temporal data in the medical context.   In the following section we review prior work on visualization of EHRs, querying of temporal data and data-driven approaches in the medical context and explain how \textit{Portinari} is positioned with respect to what has been developed. 

\subsection{Interactively Visualizing EHRs}
There has been extensive research on the development of interactive visualization tools to explore and query EHRs~\cite{LifeLines}~\cite{LifeFlow}~\cite{EventFlow}~\cite{OutFlow}~\cite{DecisionFlow}.  They aim to aid in medical decision making, and to translate the costs of having an EHR into actual benefits for physicians and patients~\cite{EhrVisSurvey}. 
Some representative systems of this approach are:\\
\noindent \textit{LifeLines}~\cite{LifeLines}, which was developed in the late 1990s and allowed the visualization of a \emph{single patient record}. The tool visualizes the timeline of cases, placements, assignments, and reviews for a patient. \textit{Portinari} in contrast aggregates patients who are similar and shows trajectories over time or events for a given variable such as diagnosis. The width of the flow in a trajectory allows rapid comparison of two trajectories at the societal level instead of an individual. Aggregation also means that individual identity will not be revealed when it comes to communicating paths taken by patients.\\
\noindent \textit{LifeFlow}~\cite{LifeFlow}, which uses a visualization inspired in an Icicle Tree to simultaneously display multiple event sequences. This is again hierarchical visualization of event sequences in the life of a unique patient. The complex visualization in the tool  does not allow exploration of trajectories of multiple patients in one visualization. \textit{LifeFlow} is useful when a patient has several exams and visits to a doctor and the health records need to be organized. It however, unlike \textit{Portinari}, does not visualize societal behaviour.\\
\noindent \textit{EventFlow}~\cite{EventFlow}, which implemented a simple query language that allows users to specify intervals between two or more events and visualize amount of flow between them. \textit{Portinari} goes one step further and visualizes alternative trajectories starting with an event sequence as an input. \\
\noindent \textit{OutFlow}~\cite{OutFlow}, which aggregates multiple event sequences into pathways, and allows users to explore external factors that influence  specific event transitions in a pathway. \textit{OutFlow} also allows clustering of events of high risk together. \textit{OutFlow} and \textit{Portinari} are very similar in terms of user experience and ease of learning. However, one of the fundamental differences between both is the query system allowing the specification of constraints on both temporal and non-temporal attributes giving  higher control over the visualization of trajectories. Moreover,  \textit{Portinari} is open source and available upon request unlike IBM's Outflow which is proprietary. \\
\noindent \textit{DecisionFlow}~\cite{DecisionFlow}, which allows the analysis of high-dimensional temporal event sequence data. It is a commercial tool that takes \textit{OutFlow} one step further to make it a full-fledged tool for ad-hoc statistics and epidemiology. \textit{Portinari} aims to be open-source alternative that integrates several statistical analysis features of \textit{DecisionFlow} along with data mining algorithms to find explanations for patterns seen in patient trajectories. \\
Rind \textit{et al.} present a comprehensive survey where some of the above interactive visualization systems are compared and evaluated~\cite{EhrVisSurvey}.

\subsection{Querying Temporal Data}
Research has also been done to develop tools and methods to allow domain experts to perform queries in temporal data, which is often too technical to be done in traditional query languages such as \textit{SQL}\cite{COQUITO}. 
A strong motivation in this context is to allow domain experts to create cohorts, groups of individuals with common features.
A representative system built for this purpose is \textit{COQUITO}~\cite{COQUITO}, a visual interface that assists the users to define cohorts with temporal constraints, giving information on the filtered population as the restrictions are added.
Systems that aim to make temporal querying  user-friendly include: 
\textit{(a)} \textit{PatternFinder}~\cite{PatternFinder}, where users can formulate queries on patient event histories with connected boxes. 
\textit{(b)} \textit{QueryMarvel}~\cite{QueryMarvel}, where a comic strip metaphor is used in order to make the querying system more easy and fun to use. 
\textit{(c)} \textit{DataPlay}~\cite{DataPlay}, which allows for a interactive trial-and-error query specification 
\textit{(d)} \emph{DataMeadow}~\cite{DataMeadow}, which provides the constructions of interactive queries through \emph{starplots}. Textual temporal query languages such as \textit{T-SQL}~\cite{TSQL} and \textit{TQuel}~\cite{TQUEL} are also relevant in our context.

As pointed by Krause \textit{et al.}\cite{COQUITO}, some of aforementioned systems allow users begin with a overview of the health data, and then filter towards a desired pattern. We may refer to this as a \textit{pattern recognition}) based approach where a user has a bird's eye view of the data and then he/she drills down in to patterns of interest.

\textit{Portinari}'s, leverages the paradigm of \emph{pattern specification} instead of \textit{pattern recognition}, by allowing specification of queries of temporal sequence of events along with personal attributes as constraints instead of filtering to a cohort from a overview of the complex dataset. Textual temporal query languages along with tools such as \textit{QueryMarvel}, \textit{DataPlay}, \emph{DataMeadow} also fall in the category of pattern specification approach.

\subsection{Data-Driven Methods in Medicine}
Our work can also be set into a larger context of applying new data-driven computational methods and tools in  medicine~\cite{PredictiveDataMining}, particularly to public health. Data-driven approaches are used to analyze data, create predictive models, help clinicians  take decisions and provide more personalized diagnosis and treatment to patients~\cite{PredictiveDataMining}~\cite{BigDataHealthCare}~\cite{MiningHealthRecords}~\cite{BigDataHealthCare2}.
Within this paradigm, unlike in traditional medical research, the data is not purposely collected to test specific questions, but obtained from legacy databases or data warehouses and used for secondary analysis~\cite{PredictiveDataMining}. 

Some representative work that fits into this scenario include the creation of disease trajectories for better understanding the correlation between diseases and their progression~\cite{DiseaseTraject},  and the system proposed by Zamora \textit{et al.} to discover and analyse co-occurrence of diagnostics, intervantions and prescriptions~\cite{Polymedication}. \textit{PARAMO}~\cite{Paramo}, is a platform developed by Ng \textit{et al.} to simplify the process of building predictive models from \textit{EHRs}.

\textit{Portinari} can be set in the broader context of incorporating data-driven methods into exploration and verification of peoples' behaviour in socio-technical systems. Portinari tries to use data that has already been collected in order to extract useful knowledge to improve public health systems. However, it differs from much of the mentioned work as it relies on a user to query the system, and does not yet create a predictive model. \textit{Portinari}'s idea is to allow epidemiologists and health care planners with a easy to use tool to test  hypotheses and find patterns or generate new hypothesis for further investigation.

\section{Conclusion}
\label{sec:conclusion}

Data in a socio-technical systems constantly evolves with how society follows new trends and changes its behavior. The effectiveness of such a system relies on how well it is able to achieve its societal goal. We believe that software verification and validation must go beyond building bug-free software to evaluating whether its societal goals are met. For instance, in cervical cancer screening the societal goal is to screen people before they get cervical cancer. This article presents \textit{Portinari}, a tool to explore and eventually verify and validate socio-technical systems developed for public health. \textit{Portinari} leverages software technology such as graph databases and web-based interactive visualization to give users insight into a societal process in real-time as data evolves. We use it to evaluate an important part of the Norwegian triage algorithm for cervical cancer screening: \emph{cytology exams to be taken at three year intervals}. We also use it to compare two different paths taken by patients who are recommend the HPV test. 

Applications of \textit{Portinari}  in the cervical cancer screening program include: \textit{(a)} Gaining insight into how effective a screening exam such as the Pap smear is in the early detection of cancer. Several,  biotechnology companies propose different laboratory testing techniques. We can use \textit{Portinari} to evaluate the long-term effectiveness of one test over another. This has several implications in terms of public spending in health. 
\textit{(b)} \textit{Portinari} can also be used by an individual in the general public to evaluate different options in waiting time before the next test is taken in order to minimize risk. However, this would require \textit{Portinari} to be adapted to different user groups ranging from patient in the general public to experienced epidemiologists. Tools like \textit{Portinari} aim to create a feedback loop between information systems run by government and the society it influences. Interactive visualizations using tools such as \textit{Portinari} can spawn self-adaptivity in the socio-technical system for cervical cancer screening if made publicly available.

\subsection{Future Work}
\label{sec:sec:fw}

During the collaborative project with the Cancer Registry of Norway we identified several areas in which software engineering research can benefit public health programs such as cervical cancer screening. We will consider the following ideas as part of future work:

\noindent \textbf{Application of software V\&V techniques:} The triage algorithm on the population is a starting point to apply software V\&V techniques to public health. The algorithm can either be represented in a declarative form (constraint satisfaction problem) or can be executed algorithmically to generate the so called \textit{ideal} patient trajectories for screening. These trajectories can be searched for in real data to validate if a trajectory recommended by the triage algorithm indeed is effective. For instance, if someone who went to screening for three consecutive invitations got cancer despite a normal cytology exam then the triage algorithm is not optimal for at least one person.  Similarly, one may also verify to what extent a patient follows a triage algorithm. The sequence of events in the life of a patient can be separated into predicates that can be verified against constraints in the triage algorithm using a formal method such as Alloy\cite{jackson2006software} or constraint logic programming~\cite{jaffar1987constraint}. 

\noindent \textbf{Data mining in socio-technical software systems:} The vast amount of data in public health often contains common patterns that serve as explanation for observed societal behavior. For instance, what is common in patients who regularly take the cytology exam? Is it their age, is it their low-risk lifestyle of  not smoking, or is it correlated with an event such as the death of some important due to cancer? Data mining tools can help extract such explanations by, for instance, finding the maximum common sequence in the temporal event sequences of thousands of people. These explanations can also be seen as a hypothesis for further statistical investigation.

\noindent \textbf{User experience studies:} \textit{Portinari} can be targeted to  epidemiologists, health policy makers, and even the general public. User experience studies will give insight into improving the interface for specific user groups. 

\noindent \textbf{Conceptual modeling of data:} Querying big data from public health can be very computationally expensive if the information we seek in the data is hard to match. In this paper, we explore the use of graph databases to transform a simple set of records to a sequence of events. Graph databases make querying very user-friendly, however, their performance to return results is still limited by the computational complexity of subgraph matching, which is NP-hard. Therefore, alternative models of data in socio-technical  systems is an active research area. We intend to explore string representations and string matching as a promising alternative for improving performance.

\section*{Acknowledgment}
We thank the Norwegian Research Council for funding our work through the Certus-SFI scheme. We also thank the Cancer Registry of Norway and the Brazilian institutions CAPES, CNPq and FAPEMIG for supporting the work.
\vspace{7mm}

\balance
\bibliographystyle{IEEEtran}
\bibliography{paper}

\end{document}